\pdfoutput=1
\documentclass[prd,aps,preprint,amsmath,nofootinbib,amssymb,eqsecnum,showkeys,tightenlines]{revtex4-1}

\usepackage{amsfonts}
\usepackage{multirow}
\usepackage{caption}
\usepackage{subcaption}
\usepackage{amsmath}
\usepackage{slashed} 
\usepackage{verbatim,graphics,graphicx,color,slashed}

\usepackage[usenames,dvipsnames]{xcolor}

\usepackage{ulem} 
\usepackage[colorlinks=true,
            linkcolor=red,
            urlcolor=blue,
            citecolor=blue]{hyperref}

\begin{document}

\title{Interference effects of two scalar boson propagators on the LHC search for the singlet 
fermion DM }

\author{P. Ko}
\email{pko@kias.re.kr}
\affiliation{School of Physics, Korea Institute for Advanced Study, Seoul 130-722, Korea}

\author{Jinmian Li}
\email{jmli@kias.re.kr}
\affiliation{School of Physics, Korea Institute for Advanced Study, Seoul 130-722, Korea}

\begin{abstract}
A gauge invariant UV-completion for singlet fermion DM interacting with the standard model (SM) 
particles involves a new singlet scalar. Therefore the model contains two scalar mediators,  mixtures of the SM 
Higgs boson and a singlet scalar boson. 
Collider phenomenology of the interference effect between these two scalar propagators
is studied in this work. 
This interference effect can be either constructive or destructive in the DM production cross section 
depending on both singlet scalar and DM masses, and it will soften the final state jets in the full mass region. 
Applying the CMS mono-jet search to our model, we find the interference effect plays a very important role in the 
DM search sensitivity, and the DM production cross section of our model is more than one order of magnitude 
below the LHC sensitivity at current stage.  
\end{abstract}
 
\maketitle

\section{Introduction}
The existence of non-baryonic Dark Matter (DM) has been established by many astrophysical observations from gravitational effects~\cite{Ade:2015xua}. However, the nature of DM and its interactions with other particles and
among themselves remain unknown. Probing the DM signals at the hadron collider could elucidate the particle physics properties of DM without suffering from astrophysical uncertainties thus becomes one of the main object 
of the current and future LHC experiments as well as future colliders such as 100 TeV $pp$ collider and high 
energy lepton colliders such as ILC or FCC-ee, CEPC. 
The strategy for DM search at colliders is rather simple: to look for events with large momentum unbalance 
that is produced by recoiling the DMs against energetic detector reconstructable objects such as jets. 

The DM effective field theory (EFT)~\cite{Goodman:2010yf,Goodman:2010ku,Duch:2014xda} which is supposed 
to be low energy approximation to a renormalizable theory for DM by integrating out the heavy particle that 
mediates the DM interaction has been widely used in early LHC analysis.  
The scalar $\times$ scalar operators describing the Dirac fermion DM $\chi$ interacting with Standard Model 
(SM) particles take the forms
\begin{align}
\frac{1}{\Lambda^2} (\bar{\chi}\chi) (\bar{f}f), ~~~ \frac{1}{\Lambda^3} (\bar{\chi} \chi) \text{Tr} (G^{\mu\nu} G_{\mu\nu}) , \label{eq:oeff}
\end{align} 
where the scale $\Lambda$ has dimension of mass, $f$ correspond to SM fermions and $G^{\mu\nu}$ 
is the gluon field. 
Then the limits on the DM relic density, DM direct/indirect detection and DM collider searches can be 
presented  in terms of a common coefficient $\Lambda$ and DM mass.   
However the EFT description might be useful only in the low energy phenomenology such as DM direct 
detection. At high energy scale such as at the LHC,  unitarity  shall be violated in general in the EFT approach 
~\cite{Buchmueller:2013dya,Busoni:2013lha,Busoni:2014sya,Busoni:2014haa}.
Also thermal relic density or indirect detection would require extensions of operator basis involving other 
SM particles such as heavy quarks ($c, b, t$, etc.), the SM Higgs boson or electroweak gauge bosons. 

In order to cure this problem of the DM EFT,   simplified model frameworks have been proposed to undertake 
the DM searches at colliders~\cite{Abdallah:2014hon,Abdallah:2015ter,Abercrombie:2015wmb}. 
The DM model Lagrangian with the minimal scalar mediator is defined as (Ref.~\cite{Buckley:2014fba}
for example) 
\begin{align}
\mathcal{L} = \mathcal{L}_{\text{SM}}  + \frac{1}{2} (\partial_\mu \phi)^2 - \frac{1}{2} m^2_\phi \phi^2 + i \bar{\chi}  \slashed{\partial} \chi - m_\chi \bar{\chi} \chi  - g_\chi \phi \bar{\chi} \chi - \sum_f g_v \frac{y_f}{\sqrt{2}} \phi \bar{f} f ~.~ \label{eq:lsimp}
\end{align}
The model has 5 new parameters, DM mass $m_\chi$, scalar mediator mass $m_\phi$, DM-mediator coupling $g_\chi$, SM-mediator coupling $g_v$ and the mediator decay wdith $\Gamma_{\phi}$,
to which the collider search bounds can be applied. 
The collider phenomenology of simplified models with different mediator spins are also studied, in gluon gluon fusion production channel~\cite{Harris:2014hga,Harris:2015kda} and in vector boson fusion production channel~\cite{Khoze:2015sra} if interactions between SM gauge bosons and the scalar are included. 
However, the simplified model Lagrangian~(\ref{eq:lsimp}) still has some problems, since  the term 
$\frac{y_f}{\sqrt{2}} \phi \bar{f} f$ violates the SM gauge symmetry explicitly 
~\cite{Talk:ko_talks,Baek:2015lna,Englert:2016joy},  because $\phi$ is a SM singlet scalar and $\bar{f} f$ is 
$SU(2)_L$ doublet. Therefore this simplified model may not be a suitable approximation/simplification of a 
UV-complete model for singlet fermion DM. 

The simplest way to write down a renormalizable and gauge invariant UV completion for effective operator~(\ref{eq:oeff}) is to introduce a singlet scalar $s$ that interacts with SM particles by the mixing with the SM Higgs boson, 
where the singlet DM can be either a fermion~\cite{Baek:2011aa} or a vector~\cite{Baek:2012se} particle.  
The LHC and the ILC searches for those minimal Higgs portal DM models with the full SM gauge symmetry are 
discussed in Ref.~\cite{Talk:ko_talks,Baek:2015lna,Bauer:2016gys} and in Ref.~\cite{Ko:2016xwd}, respectively. 
And importance of the gauge invariance in DM collider phenomenology within other setups and/or contexts 
have also been studied in Ref.~\cite{Bell:2015sza,Choudhury:2015lha,Kahlhoefer:2015bea,Ko:2016zxg,Bell:2016uhg}. 

In this work, we continue the study of collider search for singlet fermion DM with scalar mediators within the 
gauge invariant and renormalizable Higgs portal DM model discussed in Ref.~\cite{Baek:2011aa}. 
Comparing to previous studies~\cite{Talk:ko_talks,Baek:2015lna,Bauer:2016gys}, we consider the model with varying scalar decay width and resolved top quark in gluon gluon fusion production process. Moreover, the interference effects between two scalar propagators on both production cross sections and distributions of kinematic variables are discussed in detail. And their effects on the collider bounds for the singlet fermion DM searches at the LHC will be considered in terms of the CMS mono-jet/mono-V search.

This paper is organized as follows. A brief introduction on the gauge invariant UV completion of effective operator  
(1.1) is given in Sec.~\ref{sec:model} in the framework of Higgs portal DM model. In Sec.~\ref{sec:intf}, taking the gluon-gluon fusion process as a representative example, we highlight the interference effects on both DM production cross section and final state distributions. More realistic applications to a CMS DM search including all dominant DM production processes are studied in Sec.~\ref{sec:cms}.
The results are summarized in Sec.~\ref{sec:conl}.

\section{A gauge invariant model for singlet fermion DM with Higgs portal}
\label{sec:model}
In this section, we discuss the simplest Higgs portal singlet fermion DM model with SM gauge 
invariance  and renormalizability. 
The model contains a SM singlet Dirac fermion DM $\chi$ and a singlet scalar mediator $S$ in addition to the SM particles. In order to preserve the SM gauge symmetry, the $S$ only couples to SM particles through the mixing 
with the SM Higgs $h$. It is assumed that the DM $\chi$ is odd under $Z_2$ dark parity. 
Without $Z_2$  parity,  one could write a dim-4 operator, $\bar{l}_L \tilde{H} \chi$, that would make 
$\chi$ decay as forbidden.  And this $Z_2$ distinguishes the DM $\chi$ from the right-handed neutrino for neutrino 
masses and mixings.

The $Z_2$-invariant Lagrangian related to the DM sector can be written as follows~\cite{Baek:2011aa}:

\begin{align}
\mathcal{L}_{\text{new}} &= \bar{\chi} (i \slashed{\partial} -m_\chi - g_\chi S ) \chi + \frac{1}{2} \partial_\mu S \partial^\mu S - \frac{1}{2} m^2_0 S^2  \nonumber \\
         & - \lambda_{HS} H^\dagger H S^2 - \mu_0^3 S - \mu_1 S H^\dagger H -\frac{\mu_2}{3!} S^3 
         - \frac{\lambda_S}{4!} S^4~,~ \label{eq:lag}
\end{align}
where $H$ is the SM Higgs doublet.
When both scalar fields develop nonzero vacuum expectation value, $v_H$ and $v_S$, we can expand $H$ 
and $S$ as

\begin{equation}
H = 
\left( \begin{array}{c}
  G^+ \\
  \frac{1}{\sqrt{2}} (v_H + h + iG^0)
  \end{array} \right) , ~~~~~ S= v_S +s~,~
\end{equation}
where $G^+$ and $G^0$ are Goldstone bosons. Substituting into the Eq.~(\ref{eq:lag}), we can obtain the mass 
matrix for physical scalar fields $s$ and $h$. This mass matrix can be diagonalized by introducing a mixing 
angle $\alpha$, which is determined by the $\lambda$ and $\mu$ couplings in the $\mathcal{L}_{\text{new}}$. 
So two mass eigenstates $H_{1,2}$ can be expressed in terms of $h$ and $s$:

\begin{equation}
\left(  \begin{array} {c}
  h \\
  s
       \end{array} \right) = 
       \left(  \begin{array}{c c}
         \cos \alpha & \sin \alpha \\
         -\sin \alpha & \cos \alpha
         \end{array}  \right) 
         \left(
           \begin{array}{c}
           H_1 \\
           H2
           \end{array}
         \right)~.~ \label{eq:hmix}
\end{equation} 

We will take the $H_1$ as the SM-like Higgs that is consistent with the LHC discovery~\cite{Aad:2012tfa,Chatrchyan:2012xdj}.  So far the measurements of the 125 GeV Higgs boson signal strengths~\cite{Khachatryan:2014jba,Aad:2015gba} indicate that the mixing angle $\alpha$ is small $\sin\alpha \lesssim 0.4$~\cite{Robens:2015gla,Cheung:2015dta,Dupuis:2016fda}.

The  Lagrangian describing the interactions of  $H_{1,2}$ with the SM particles and the singlet fermion 
DM  is given by

\begin{align}
\mathcal{L}_{\text{int}} &= -(H_1 \cos \alpha + H_2 \sin \alpha) \left[ \sum_f \frac{m_f}{v_h} \bar{f} f 
- \frac{2m^2_W}{v_h} W^+_\mu W^{-\mu} - \frac{m_Z^2}{v_h} Z_\mu Z^\mu  \right] \nonumber \\ 
     & + g_\chi (H_1 \sin \alpha - H_2 \cos \alpha) ~ \bar{\chi} \chi ~.~ \label{eq:lint}
\end{align}
in the mass eigenstates for the neutral scalar bosons. 

There are four free parameters relevant to our phenomenological study,  $m_\chi$, $m_{H_2}$, $g_\chi$ 
and $\sin \alpha$. 
Moreover, there could be new decay modes open for the $H_2$, such as $H_2 \to H_1 H_1$ if kinematically 
allowed. And we  will see later that the total decay width of $H_2$ ($\Gamma_{H_2}$) is playing a very important 
role in reconstructing observables at detector  level through interference between $H_1$ and 
$H_2$ propagators.   Therefore  $\Gamma_{H_2}$  will be treated as the fifth free parameter in the following discussion. 

\section{The interference effect between two scalar  mediators at LHC}
\label{sec:intf}
\begin{figure}[htb] 
\includegraphics[width=0.32\textwidth]{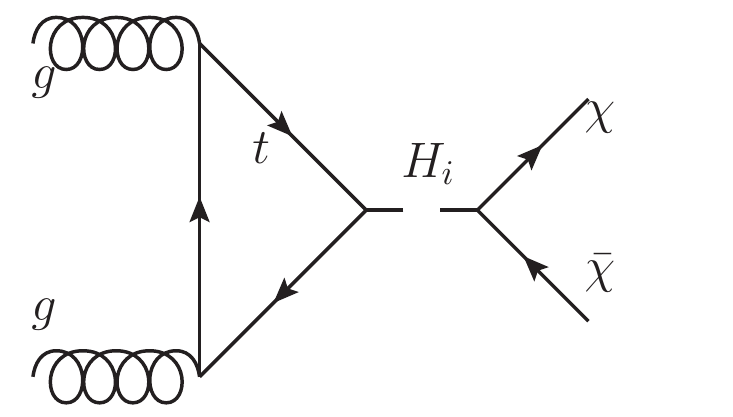}
\includegraphics[width=0.32\textwidth]{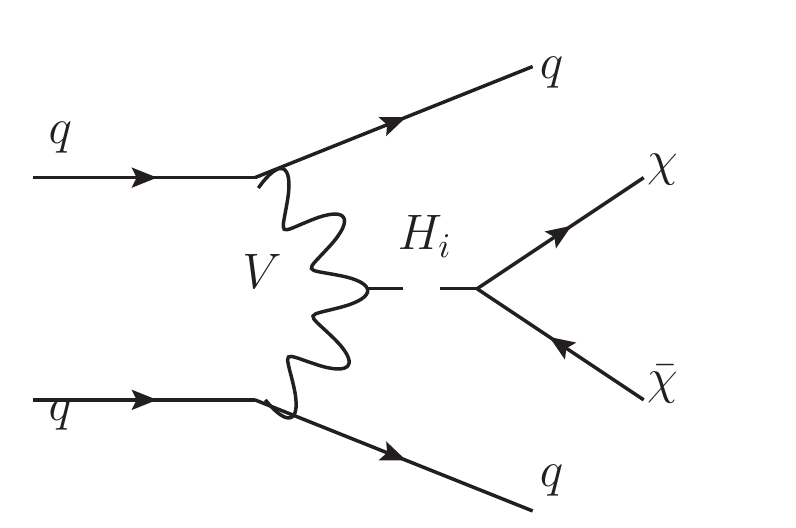}
\includegraphics[width=0.32\textwidth]{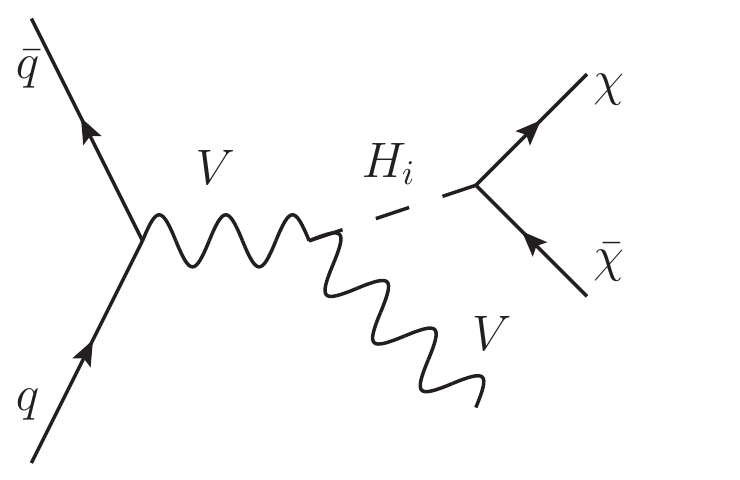}
 \caption{\label{fig:diag} The dominant DM production processes at LHC. }
\end{figure}

In the singlet fermion DM model with Higgs portal described in the previous section, the DM production is 
dominated by three processes as shown in Fig.~\ref{fig:diag}: i.e. gluon-gluon fusion (ggF), vector boson fusion 
(VBF) and Higgs Strahlung (VH).  

In contrast to the simplified scalar mediated DM model recommended by the LHC Dark Matter Forum~\cite{Abercrombie:2015wmb}, there are two propagators ($H_1$ and $H_2$) that can mediate the DM 
pair production in the gauge invariant model descried in the previous section. Note that the Lagrangian in 
Eq.~(\ref{eq:lint}) resembles the singlet scalar mediated DM model in Ref.~\cite{Abercrombie:2015wmb} when 
only fermionic couplings of $H_2$ are concerned.  

The interference between two propagators in the differential  production cross sections of the DM pair take  
the following form: 

\begin{align}
\frac{d \sigma_i}{d m_{\chi\chi}} \propto | \frac{\sin 2\alpha ~ g_\chi}{m_{\chi\chi}^2 - m^2_{H_1} + i m_{H_1} \Gamma_{H_1}} - \frac{\sin 2\alpha ~ g_\chi}{m_{\chi\chi}^2 - m^2_{H_2} + i m_{H_2} \Gamma_{H_2} }  |^2  ~,~ \label{eq:intf}
\end{align}
where $\sigma_i$ corresponds to the cross section of given production process and  $m_{\chi\chi}$ 
is the invariant mass of DM pair. The minus sign between two propagators comes from the $SO(2)$ nature 
of the  mixing matrix in Eq.~(\ref{eq:hmix}), which is found to be helpful to evade the DM direct detection
~\cite{Baek:2011aa,Duerr:2016tmh} in such class of models.  The interference effect will not only influence 
the total production rate of DM pair, but also change the shape of kinematic variables. 

To give more concrete examples on the interference effect, a few assumptions are made to narrow down the parameter space. 
We will fix $\sin \alpha =0.2$ and $g_\chi=1$ in our following discussion. Because the differential cross section 
are universally proportional to $g_\chi  \sin 2\alpha $ as shown in Eq.~(\ref{eq:intf}), changing the $\sin\alpha$ 
and $g_\chi$ will simply rescale the differential cross section as long as the $\Gamma_{H_i}$ does not differ much. 
The scalar $H_1$ is identified as the 125 GeV Higgs boson with properties that are consistent with the LHC 
discovery, so that $m_{H_1}=125$ GeV and $\Gamma_{H_1} = \cos^2 \alpha \cdot \Gamma_{h_{\text{SM}}}$.
Models with $m_\chi < m_{h_{\text{SM}}}/2$ will be highly constrained by the Higgs invisible decay search at 
LHC. This usually requires very small $g_\chi$, e.g. for $\sin \alpha =0.2$, $g_\chi$ should be smaller than 
$\lesssim 0.1$ in order to satisfy the current upper bound on the invisible Higgs branching ratio: 
Br($h_{\text{SM}} \to \chi \chi$)$<0.24$~\cite{CMS:2016rfr}.  Then the DM production cross section should be 
small in such cases. The same situation exists when DM is heavy.  So we will focus on the scenarios with 
medium DM mass in this work,  which we choose $m_\chi =80$ GeV without lose of generality. 
Then we are left with two most relevant parameters: $m_{H_2}$ and $\Gamma_{H_2}$. 
 
The FeynRules~\cite{Alloul:2013bka}/MadGraph5\_aMC@NLO~\cite{Alwall:2014hca} framework is used  
in order to calculate the NLO QCD cross sections and simulate the events. The FeynRules takes the Lagrangian 
of the simplified model in Eq.~(\ref{eq:lint}) as well as the UV/$R_2$ counterterms for the NLO QCD computations 
from NLOCT~\cite{Degrande:2014vpa}/FeynArts~\cite{Hahn:2000kx} to generate the Universal FeynRules Output 
model files. The MadGraph5\_aMC@NLO uses the model files to compute the tree-level and loop-level amplitudes 
for any processes of the model. 

\begin{figure}[htb] \centering
\includegraphics[width=0.7\textwidth]{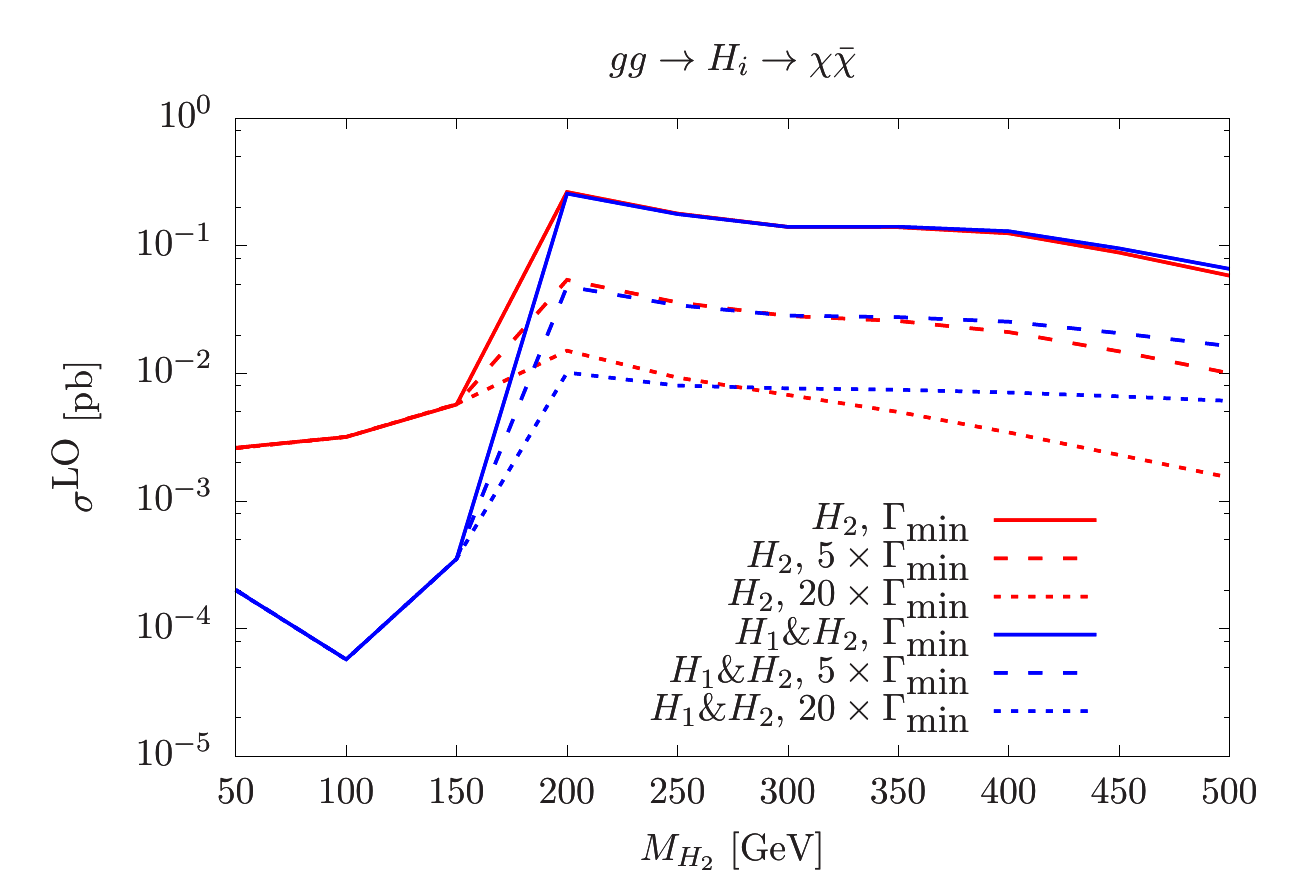}
 \caption{\label{fig:xsec} The LO cross section for gluon-gluon fusion process at 13 TeV LHC.  
 The meanings of the different line types are explained in the text and the similar strategy 
 will be used in all figures.}
\end{figure}

We calculate the Leading-Order (LO) cross section of the gluon-gluon fusion DM pair production by using the loop induced mode~\cite{Hirschi:2015iia} of MadGraph5\_aMC@NLO. The results for varying $m_{H_2}$ and $\Gamma_{H_2}$ are shown in Fig.~\ref{fig:xsec}. 

In the figure, the  $\Gamma_{\min}$ for $H_2$ is calculated by assuming $H_2$  decays only into SM particles 
and DM pair through the interactions given in Eq.~(\ref{eq:lint}), where we have set $\sin \alpha =0.2$ and $g_\chi=1$.  
Note that the actual $H_2$ decay width could be larger than $\Gamma_{\min}$, 
if $H_2 \rightarrow H_1 H_1$ is open and non-negligible, or if there are other decay channels  of $H_2$.  
For example, there could be  extra dark sector particles such as dark Higgs or dark gauge bosons if $Z_2$ 
symmetry is replaced by dark gauge symmetry (see Refs.~\cite{Baek:2014kna,Ko:2016zid} for example).  
These extra channels are more model dependent though. 
Therefore we consider three different widths of $H_2$ throughout the work: $\Gamma_{\min}$, $5\times\Gamma_{\min}$ and $20\times\Gamma_{\min}$, respectively.
The lines associate to $H_1\&H_2$ and $H_2$ are calculated with and without the $H_1$ as the mediator respectively. The former case corresponds to the gauge invariant singlet fermion DM models with 
Higgs portal, while the later case corresponds to the  usual singlet scalar portal DM model as proposed in 
Ref.~\cite{Abercrombie:2015wmb} and widely used in literature. 

From Fig.~\ref{fig:xsec}, we can observe that including the $H_1$ will substantially reduce the DM pair production 
cross section when $m_{H_2}\lesssim 2 m_{\chi}$. This is because of the destructive interference between two mediators caused by the minus sign in Eq.~(\ref{eq:intf}).  Note that the collider signatures in this 
parameter region have not been studied carefully in previous studies of the singlet fermion DM model with Higgs 
portal except in Ref.~\cite{Baek:2015lna}, partly because the signal cross section is expected to be small.  
Our study in the present work shows 
that the signal cross section is even smaller than that of the simplified model with a single
scalar mediator which is violating the full SM gauge invariance.

Once the $m_{H_2} \gtrsim 2 m_\chi $, the cross section increases dramatically due to resonant enhancement~\footnote{In this mass region, the $H_1\&H_2$ scenario can be effectively described by the $H_2$ scenario only when the decay width of $H_2$ is narrow and the mass of $H_2$ is relatively light.}. 
From Eq.~\ref{eq:intf}, we know the contributions of two propagators interfering constructively in the region $m_{\chi \chi} \in [2m_\chi, m_{H_2})$ and destructively in the region $m_{\chi \chi} \in (m_{H_2}, +\infty)$. 
When $m_{H_2}$ is not much larger than twice of DM mass, the destructive effect dominates. 
As the $H_2$ becomes heavier ($\gtrsim 270$ GeV in our parameter setup), there are more fraction of events 
falling into the constructive region. This explains why the $H_1\&H_2$ scenario has smaller cross section than 
the usual $H_2$ scenario when $m_{H_2} \in (2 m_{\chi\chi}, \sim 270~\text{GeV})$ and larger cross section 
when $m_{H_2} \gtrsim 270$ GeV. 
Such features will become even more significant for wider decay width of $H_2$ as we can expect. 
Given $m_{H_2}=400$ GeV as an example, the difference in total cross section is $\frac{\sigma(H_1\&H_2)-\sigma(H_2)}{\sigma(H_2)} \sim 4\%$ for $\Gamma_{H_2}= \Gamma_{\min}$ while can be as large as $\sim 106\%$ for $\Gamma_{H_2}=20\times \Gamma_{\min}$. 

\begin{figure}[htb] \centering
\includegraphics[width=0.47\textwidth]{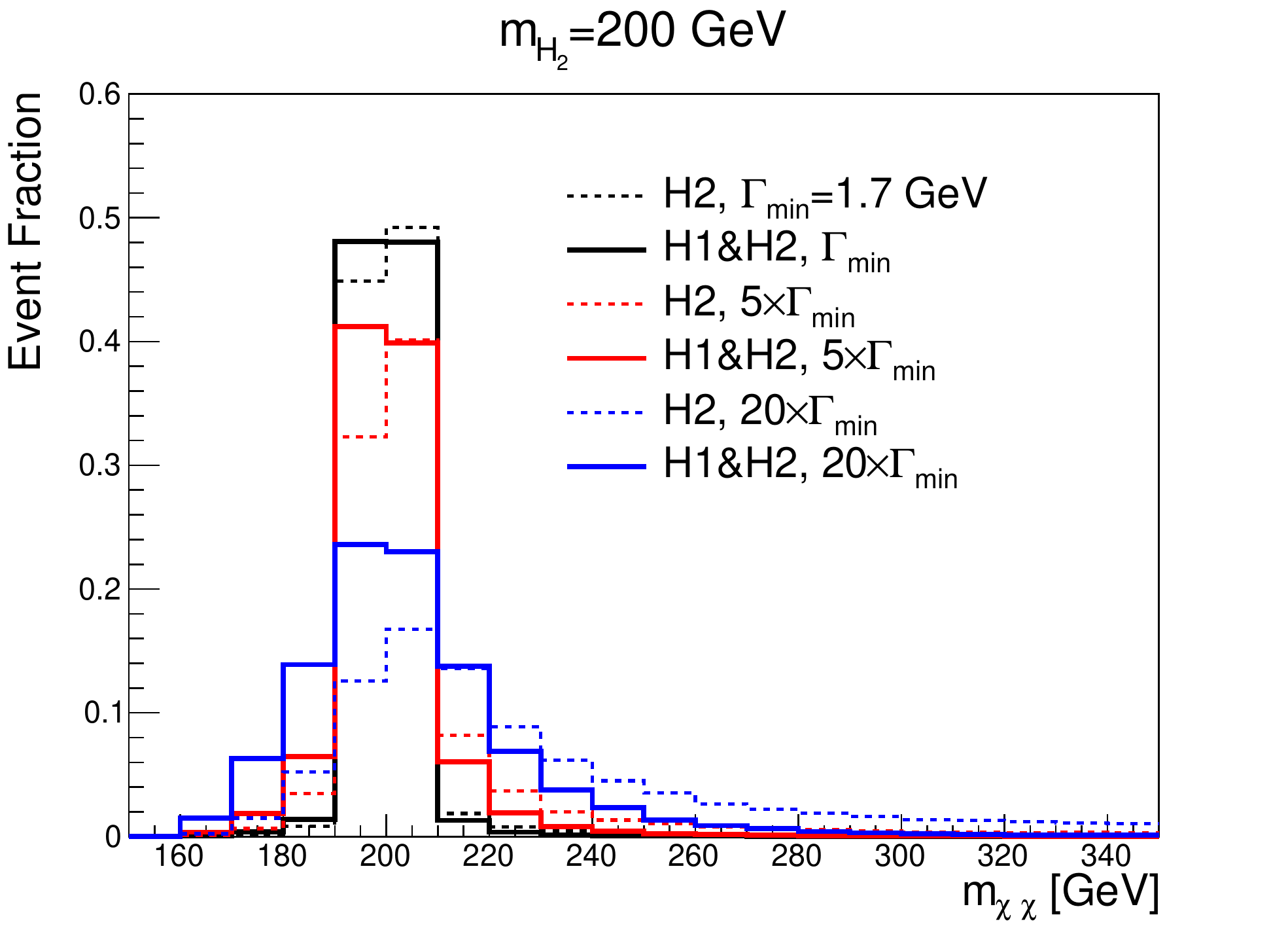}
\includegraphics[width=0.47\textwidth]{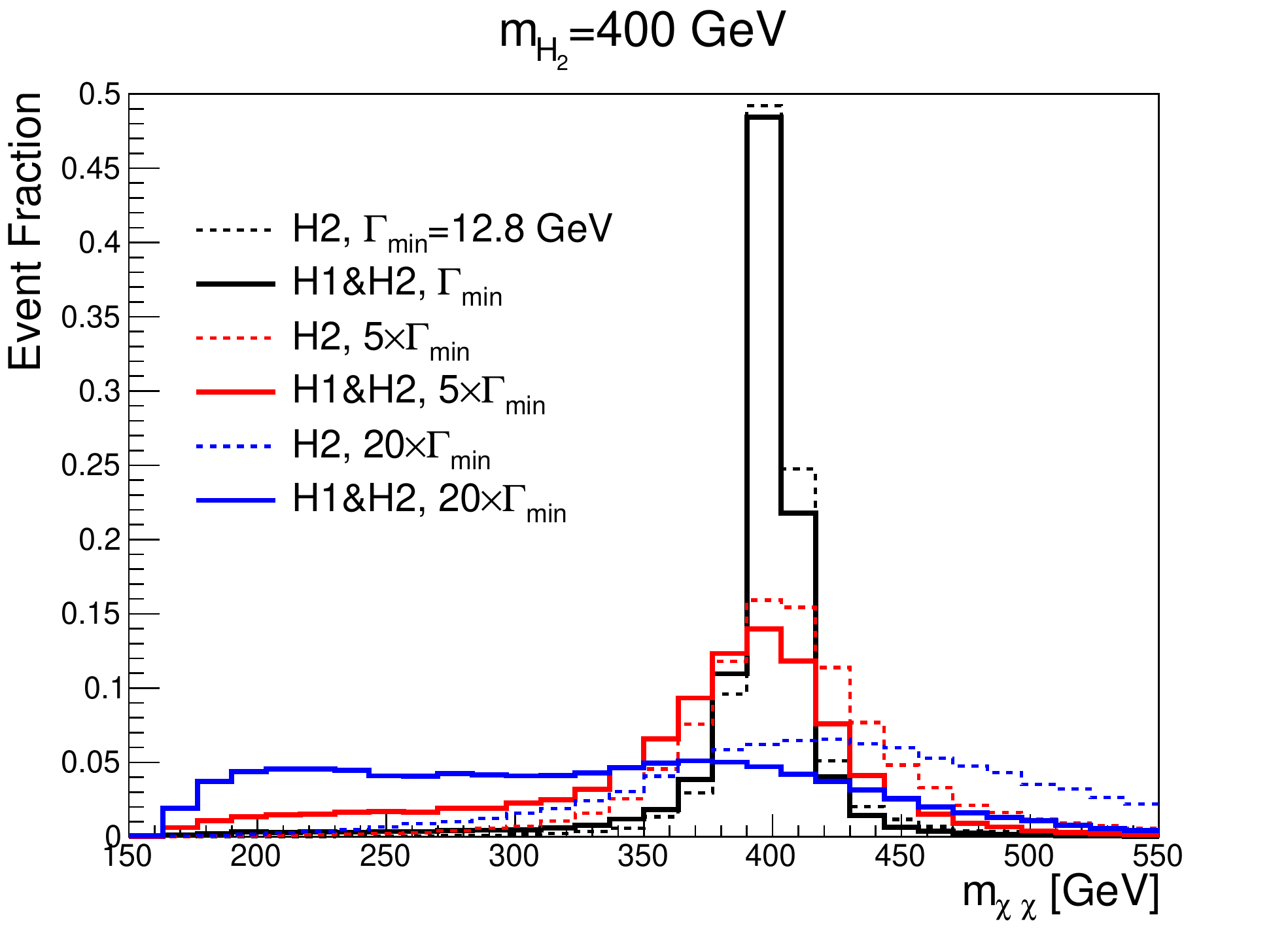}
 \caption{\label{fig:mxx} The parton level distributions of $m_{\chi\bar{\chi}}$ for gluon-gluon fusion process at 13 TeV LHC. }
\end{figure}

To see the interference effect more explicitly, we plot the differential cross section in $m_{\chi\chi}$ 
for the ggF process in Fig.~\ref{fig:mxx}. Two different masses of $H_2$ are considered with the DM 
mass being fixed to 80 GeV. For both masses, we can observe the enhancement in event fraction for 
$m_{\chi \chi} \in (2m_\chi, m_{H_2})$ and deduction in event fraction when $m_{\chi \chi} > m_{H_2}$. 
For $m_{H_2}=200$ GeV, the total event fraction in $(2m_\chi, m_{H_2})$ is smaller than that in $(m_{H_2}, +\infty)$ while it is opposite for $m_{H_2}=400$ GeV. 
Note that for heavy mass and large decay width of $H_2$, the resonant peak can be smeared out as shown 
by the solid blue curve in the right panel  due to the significant enhancement from the interference effect between $H_1$ and $H_2$.

So far, we have discussed the features of the interference effect at parton level for $gg \to H_i \to \chi \chi$. 
In practice, there will be extra radiations from the gluons in initial state and top quarks in the loop, which may 
affect the features of interference to some extent. Moreover, the variable $m_{\chi \bar{\chi}}$ is not a physical 
observable at hadron colliders since it cannot be reconstructed there. 
In order to obtain a more realistic results on the interference effect to a realistic DM search at hadron collider, 
we need to include those radiations in our events simulation and present the result with a more realistic and 
measurable  observable.  

In a typical DM search at the LHC, one usually requires an energetic jet in the final state, i.e. mono-jet search. 
So we generate the DM pair production in associate with an extra jet at parton level. The events are passed to Pythia6~\cite{Sjostrand:2006za} for parton showering and hadronization. The final state particles are used for reconstructing the physical objects such as isolated lepton and jets. Jets are clustered using the anti-$k_t$ algorithm~\cite{Cacciari:2008gp} with distance parameter of $R=0.4$ as implemented in Fastjet
~\cite{Cacciari:2011ma}.   The benchmark points with $m_{H_2}=400$ GeV  and three different decay widths 
for $H_2$ are chosen for representative study because the interference effect here is relatively large.

\begin{figure}[htb] \centering
\includegraphics[width=0.47\textwidth]{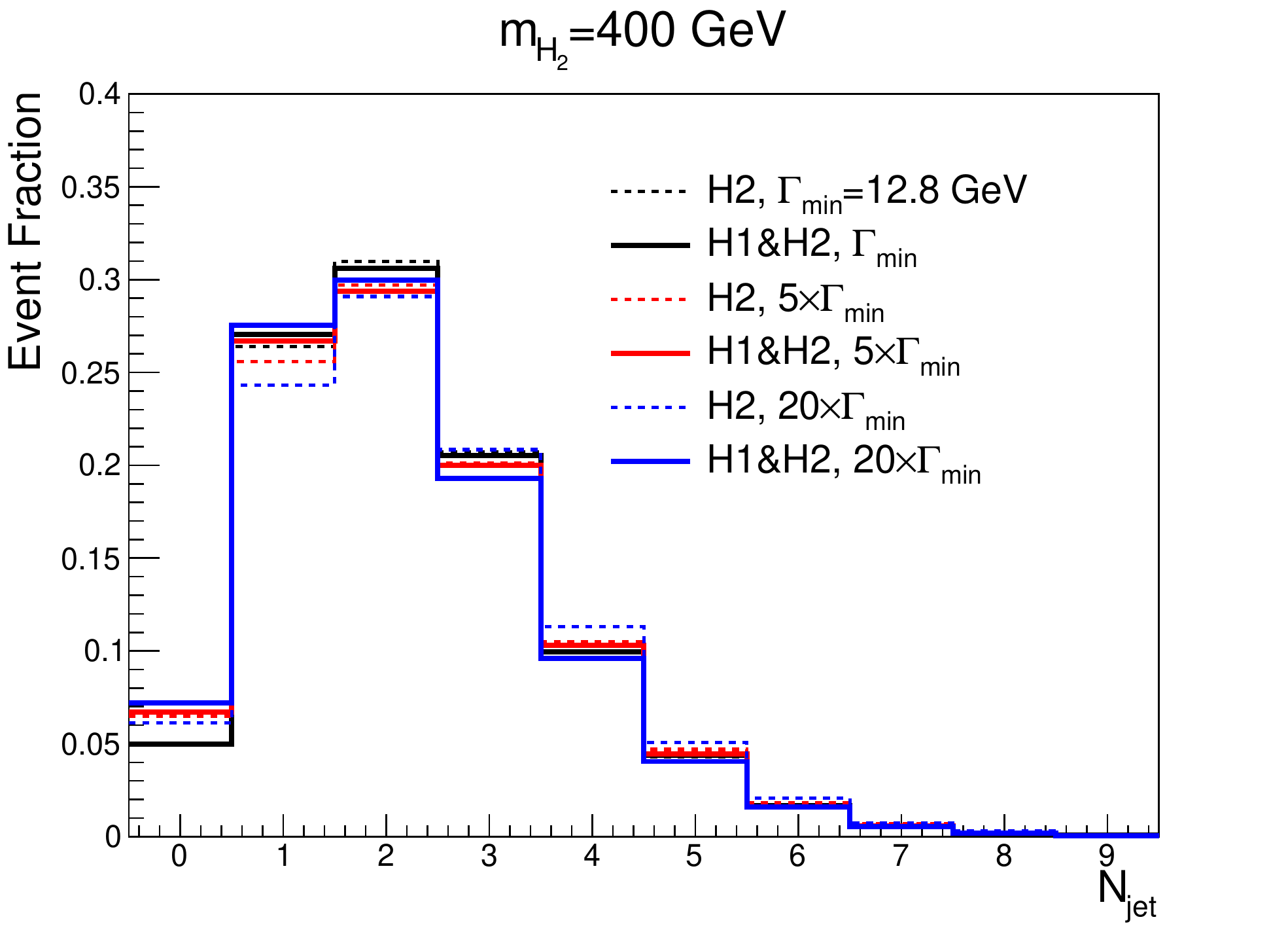}
\includegraphics[width=0.47\textwidth]{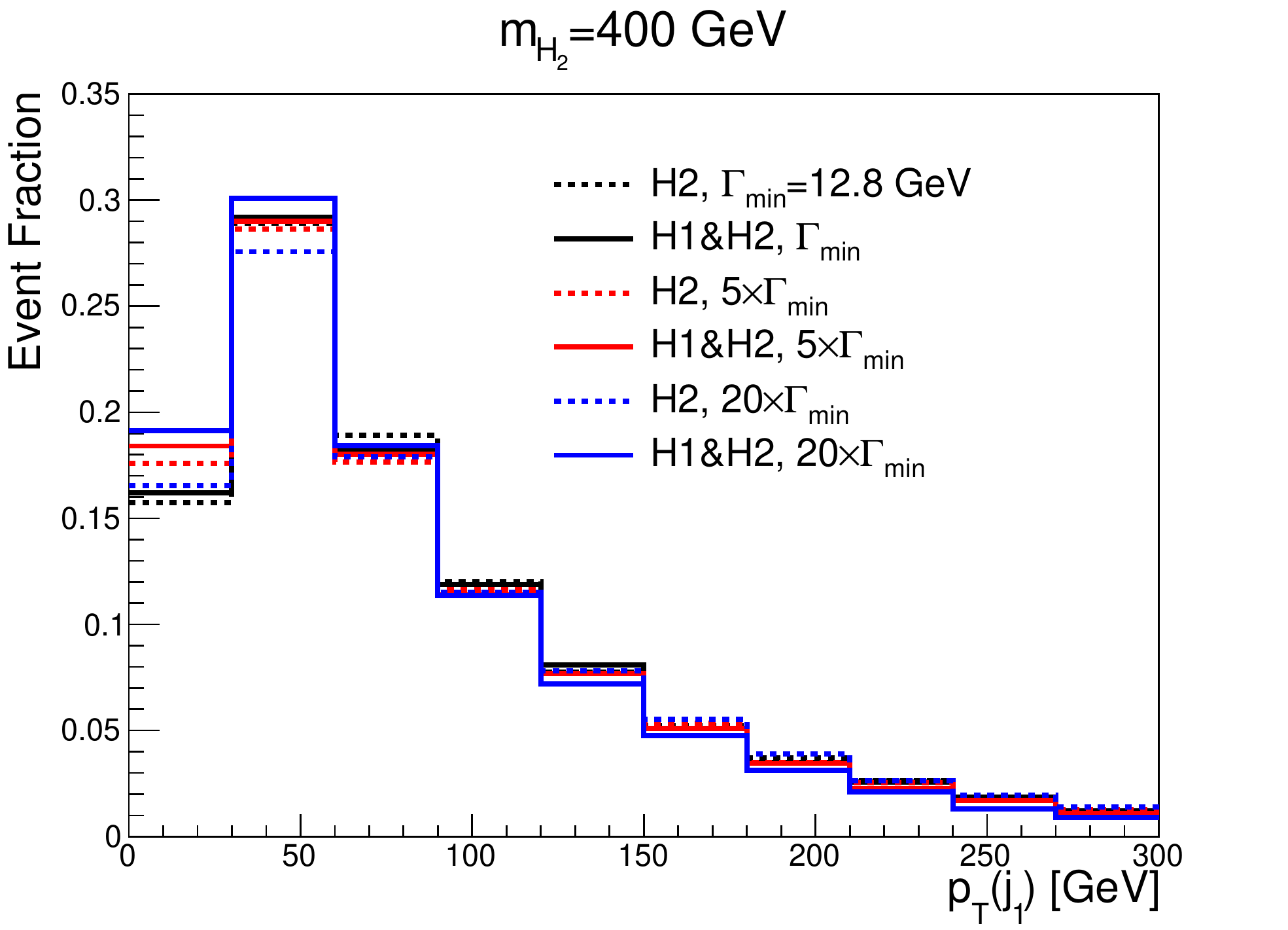}
 \caption{\label{fig:pythia} The distributions of the number of reconstructed jets (left) and the transverse momentum of the leading jet (right) for ggF process at 13 TeV LHC, including the radiation of an extra jet and parton shower effect.}
\end{figure}

The distributions of reconstructed jet number $N_{\text{jet}}$ and leading jet transverse momentum $p_T(j_1)$ are shown in Fig.~\ref{fig:pythia}. The jets are required to have $p_T(j)>20$ GeV and $|\eta(j)|<4.5$. 
Both the frequency and the energy of jet radiation are proportional to the energy scale of a event, which is given by $m_{\chi\chi}$. And the 
interference effect tends to increase the event rate in low $m_{\chi\chi}$ region and reduce the event rate in high $m_{\chi\chi}$ region. As a result, we can observe that the $H_1\&H_2$ scenario has lower jet multiplicity and 
softer $p_T(j_1)$ distribution comparing to the $H_2$ scenarios. The differences become larger for wider 
$H_2$ decay width.

\begin{figure}[htb] \centering
\includegraphics[width=0.7\textwidth]{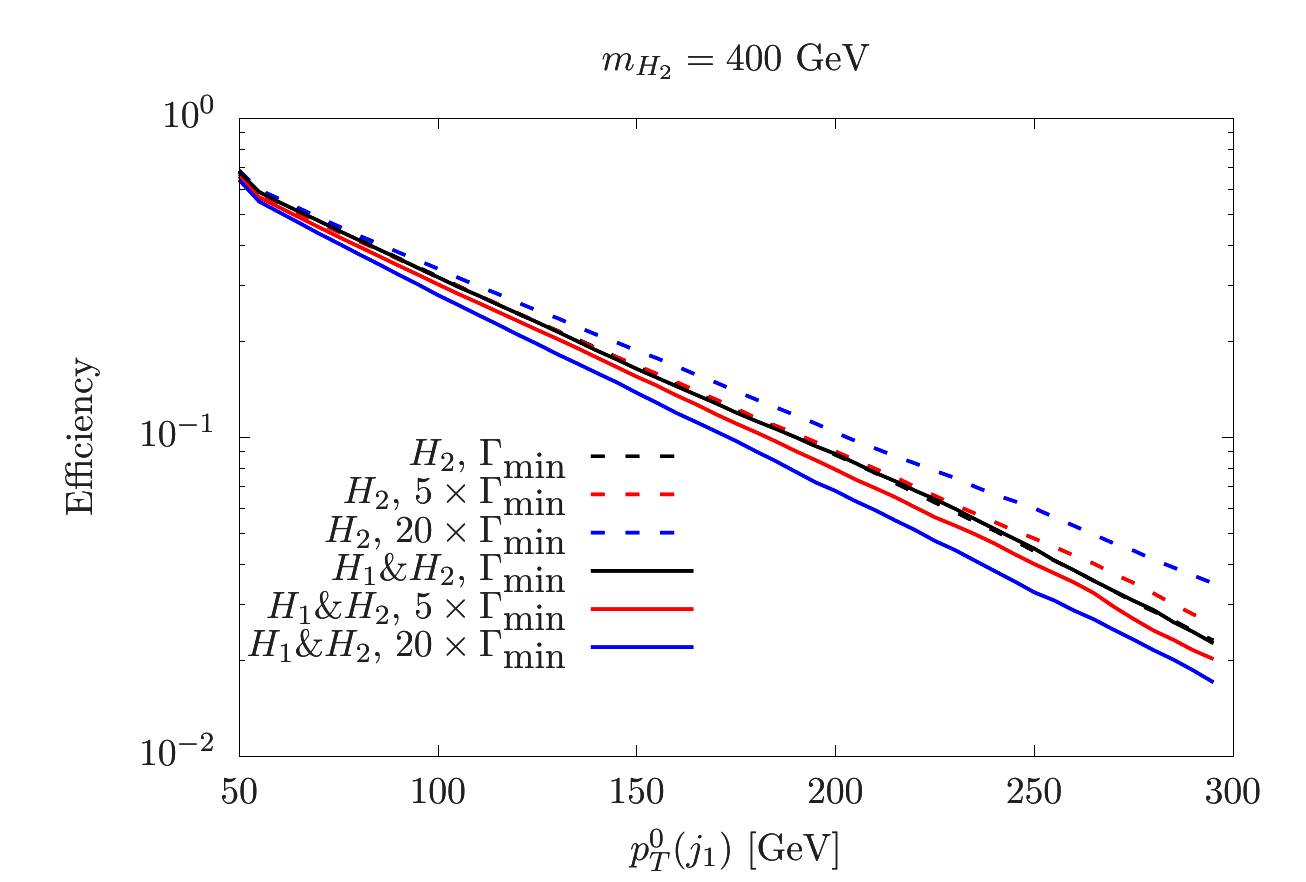}
 \caption{\label{fig:cuml} The cumulative curve of the $p_T(j_1)$ distribution in ggF process at 13 TeV LHC. }
\end{figure}

A DM search at the LHC usually counts the number of events that pass certain cuts, e.g. $p_T(j_1)>200$ GeV. 
Thus the efficiency of the cut is directly related to the search sensitivity. The efficiency for a given cut $p_T(j_1) > 
p_T^0(j_1)$ is calculated by integrating the $p_T(j_1)$ distribution in the right panel of Fig.~\ref{fig:pythia} from $p_T^0(j_1)$ to infinity, i.e. the cumulative curve. We present the cumulative curves for different scenarios in Fig.~\ref{fig:cuml}. The difference in efficiencies between $H_1\&H_2$ and $H_2$ scenarios becomes more significant for more stringent $p_T(j_1)$ cut and/or larger $H_2$ decay width. 
For example, when $H_2$ width is large $20\times \Gamma_{\min}$, the efficiency ratios (defined as 
$\frac{\epsilon(H_1\&H_2)}{\epsilon(H_2)}$) are 0.83 and 0.66 for $p_T(j_1)>100$ GeV and $p_T(j_1)>200$ GeV 
cuts, respectively. As for narrower $H_2$ width $5\times \Gamma_{\min}$ the corresponding efficiency ratios are 
0.95 and 0.88. 

\section{The LHC search bounds}
\label{sec:cms}
Motivated by the gauge invariance of simplified Higgs portal DM models, we have found that the existence 
of the additional SM-like Higgs $H_1$ propagator could affect the LHC DM search, both on total production 
rate and shape of kinematic variables.  
In this section, we will demonstrate the importance of  this influence in a practical analysis at the LHC. 

In Ref.~\cite{CMS-PAS-EXO-16-037}, the CMS collaboration reported DM searches in final states with either 
an energetic jet or a boosted hadronically decaying vector boson using 12.9 fb$^{-1}$ data set at 13 TeV.  
The search is especially relevant to our simplified model in which two Higgs bosons couple strongly to top 
quark and vector boson. 
In their search, two classes of cuts are designed aiming for the energetic jet and boosted boson respectively. 
\begin{itemize}
\item{Mono-jet cuts:} (1). Events are required to have missing transverse momentum $p^{\text{miss}}_T >200$ GeV.  (2). A event is vetoed if it contains any isolated leptons, isolated photons, $\tau$-tagged jets and b-tagged jets. (3). The jets are clustered using anti-$k_t$ algorithm with $R=0.4$ (denoted by $j^{\text{ak4}}$). The leading $j^{\text{ak4}}$ is required to have $p_T(j^{\text{ak4}}_1) > 100$ GeV and $|\eta(j^{\text{ak4}}_1)|<2.5$. (4). The minimum azimuthal angle between the $\vec{p}_T^{\text{miss}}$ and leading four $j^{\text{ak4}}$s with $p_T>30$ GeV is required to be greater than 0.5.  
\item{Mono-V cuts:} (1). A more stringent cut on $p^{\text{miss}}_T$ is applied, $p^{\text{miss}}_T >250$ GeV. (2). The final states particles are reclustered using anti-$k_t$ algorithm with $R=0.8$, denoted by $j^{\text{ak8}}$. The leading $j^{\text{ak8}}$ should has $p_T(j^{\text{ak4}}_1) > 250$ GeV and $|\eta(j^{\text{ak8}}_1)|<2.4$. (3). Invariant mass of the leading $j^{\text{ak8}}$ after pruning~\cite{Ellis:2009me} is required to be between 65 and 105 GeV. (4). The N-subjettiness variable $\tau_{N}$~\cite{Thaler:2010tr} is used to discriminate the two prong decays of the vector boson from QCD jets. The leading $j^{\text{ak8}}$ is required to have $\tau_2/\tau_1 <0.6$. 
\end{itemize}
Two signal regions (SR) are defined in their analysis based on above cuts: the mono-jet SR and mono-V SR. Events that pass both mono-jet cuts and mono-V cuts are assigned to the mono-V SR. And those that pass the mono-jet cuts while fails any of these mono-V cuts are assigned to the mono-jet SR. 
By recasting their analysis on the SM Higgs invisible decay and comparing with their results given in the Fig.15 of Ref.~\cite{CMS-PAS-EXO-16-037}, we can find that the 95\% CL expected upper limit on the number of new physics events in mono-jet SR ($N^{\text{upper}}_{\text{mono-jet}}$) and mono-V SR ($N^{\text{upper}}_{\text{mono-V}}$) are around 10833 and 447.2, respectively. The upper bound will be projected to our simplified Higgs portal DM model with either one or two scalar bosons. 

The DM pair can be produced by three processes in our model as shown in Fig.~\ref{fig:diag}.
The ggF process itself does not produce any observable signals at detector. Extra energetic jets radiation from either initial state gluon or top quark in the loop can circumvent this issue. The LO cross section for DM pair production in association with a jet is computed by the MadGraph5\_aMC@NLO, where the jet is required to have $p_T(j)>100$ GeV. Meanwhile, the higher order corrections are found to be quite significant in improving the ggF cross section in Higgs production. Using the SusHi program~\cite{Harlander:2012pb,Harlander:2016hcx,Harlander:2002wh}, the NNLO K-factors for Higgs mass $\in [100,500]$ are calculated to be around 2.5. So the production cross section for the ggF process is given by multiplying the LO cross section in MadGraph5\_aMC@NLO with a universal K-factor of 2.5. 
Two forward/backward jets are produced in associate with DM pair in VBF processes. The correction to the inclusive VBF Higgs production cross section up to NNLO is found to be only around percent level~\cite{Cacciari:2015jma}. So the LO cross section for DM production in VBF process can be used directly, where we only impose mild cuts on forward/backward jets, $p_T(j)>20$ GeV and $|\eta(j)|<4.5$. 
As for the VH process, we include all three vector bosons $W^\pm$ and $Z$ in the final state. Even though only hadronically decaying Vs are considered in the CMS analysis, we do not include the decay branching ratio of the vector bosons when calculating the DM production cross section. 
The NLO cross section in QCD is calculated by the MadGraph5\_aMC@NLO. 

\begin{figure}[htb] \centering
\includegraphics[width=0.47\textwidth]{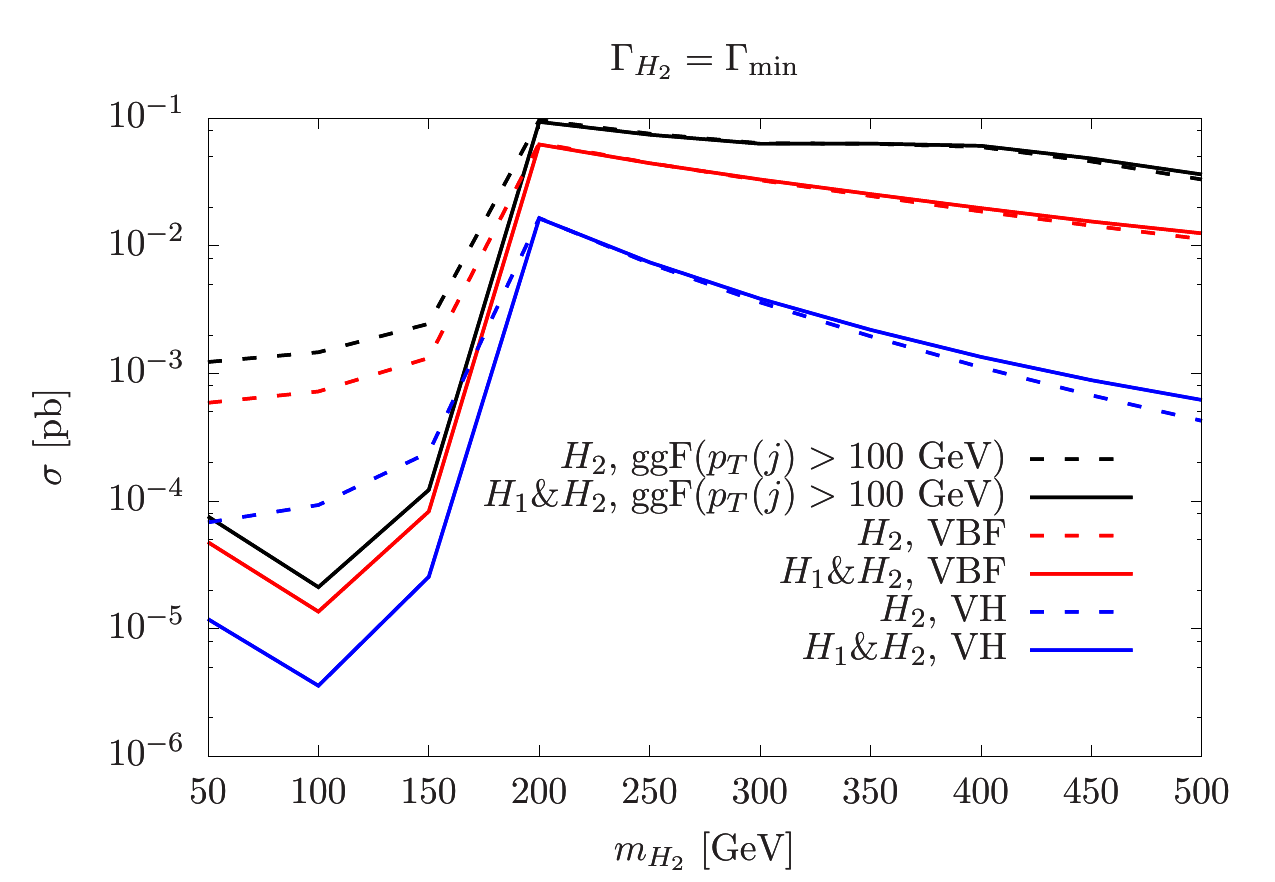}
\includegraphics[width=0.47\textwidth]{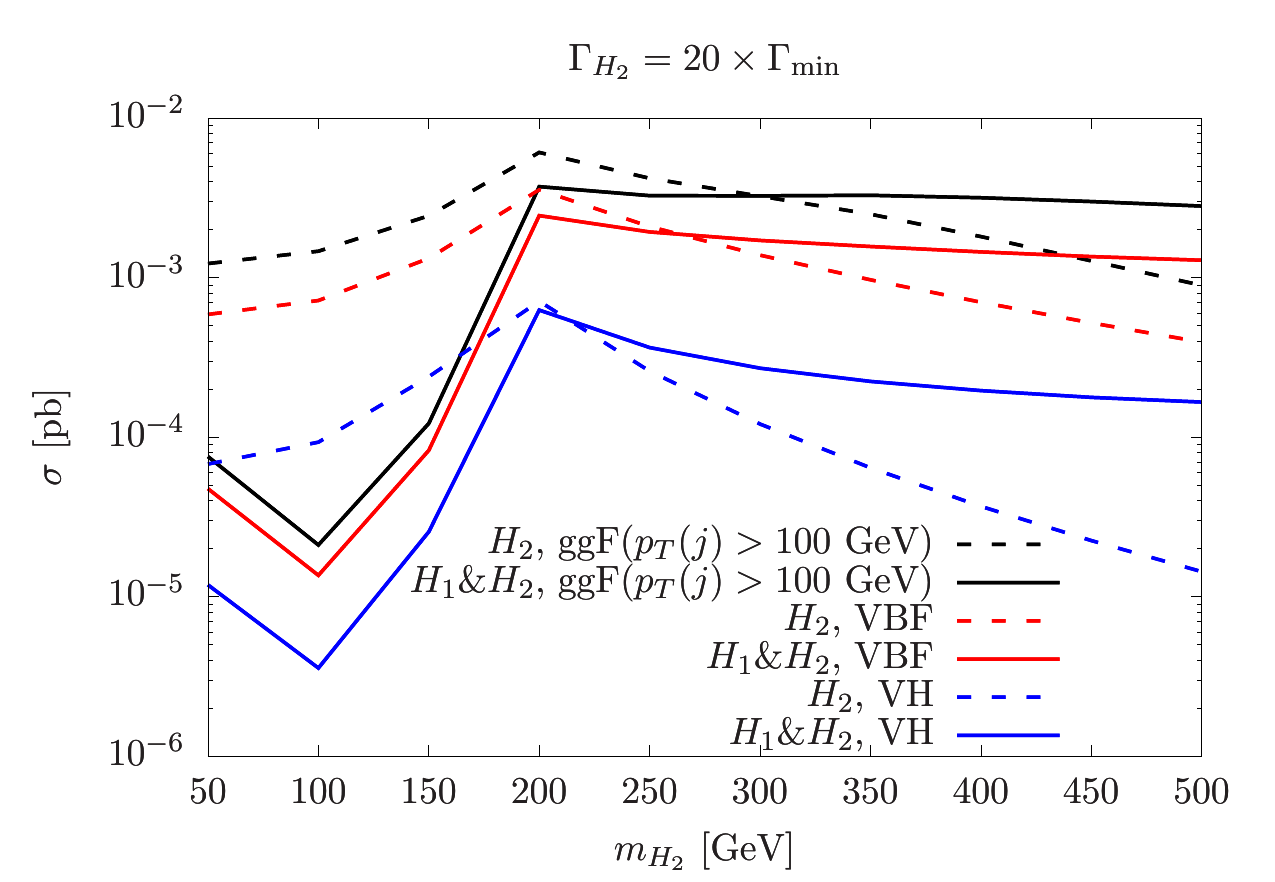}
 \caption{\label{fig:xsec3} The production cross sections for ggF (black), VBF (red) and VH (blue) processes at 13 TeV LHC. Left panel: The decay width of $H_2$ is $\Gamma_{\min}$. Right panel: A relatively large decay width of $H_2$ is chosen, $20\times \Gamma_{\min}$. 
 The difference between the solid and the dashed lines with the same 
 color show the importance of the interference effects from two scalar boson propagators.}
\end{figure}

In Fig~\ref{fig:xsec3}, we plot the cross sections of three DM production processes at the 13 TeV LHC. 
The main features of this figure regarding the variations of $m_{H_2}$ and $\Gamma(H_2)$ follow the general 
arguments that are conducted for Fig.~\ref{fig:xsec}. Beside those, we can observe that the ggF is the most 
dominant production process even after applying the stringent cut on the radiated jet. The cross section of VH is around 
one to two orders of magnitude smaller than that of ggF. Among three DM production processes, the interference 
effect between two propagators is most significant in VH process. This is because the VH process has higher 
energy scale than ggF/VBF process due to the heavy vector boson in the final state. The fact that the parton 
distribution function of proton favoring small $x$ (which is the energy fraction for a given parton inside the proton) helps to increase the production probability in the low energy scale region, i.e. the constructive interference region. 

Next, we should simulate events for calculation of the signal efficiencies in mono-jet search and mono-V search. Signal events are simulated based on the FeynRules/MadGraph5\_aMC@NLO framework as introduced before.  The Pythia6 is used for decaying the vector boson in VH process, as well as for parton showering and hadronization. The realistic detector effects are simulated by the Delphes3~\cite{deFavereau:2013fsa}. For each production process, the same cuts that used in computing the cross section are used in event generation. 

\begin{figure}[htb] \centering
\includegraphics[width=0.47\textwidth]{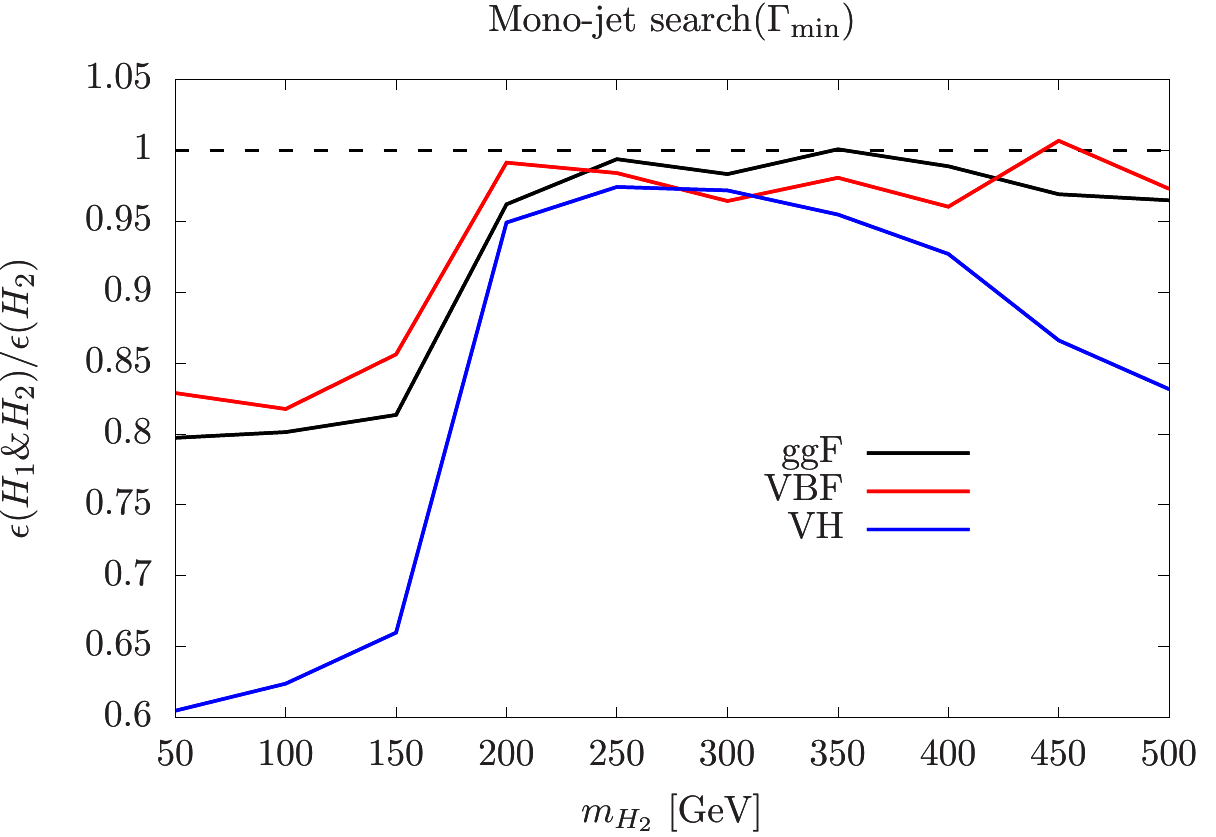}
\includegraphics[width=0.47\textwidth]{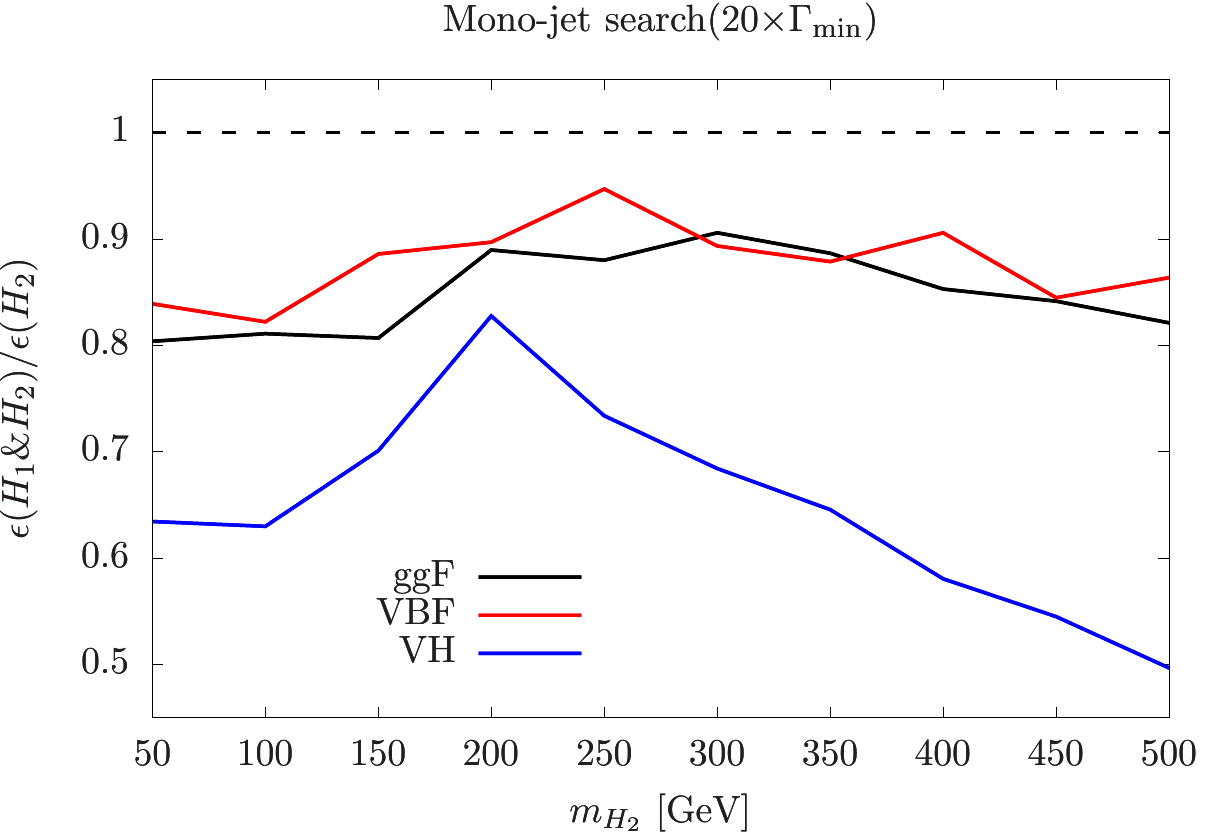}\\
\includegraphics[width=0.47\textwidth]{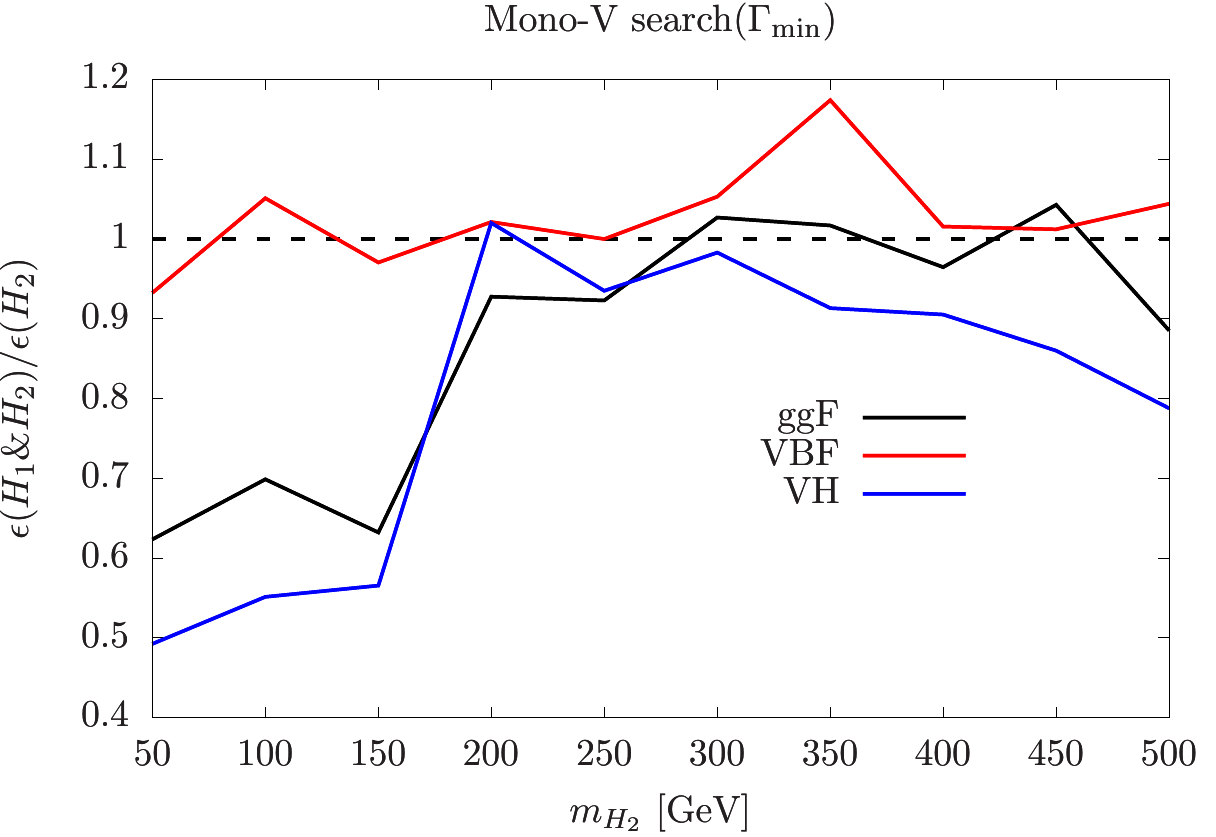}
\includegraphics[width=0.47\textwidth]{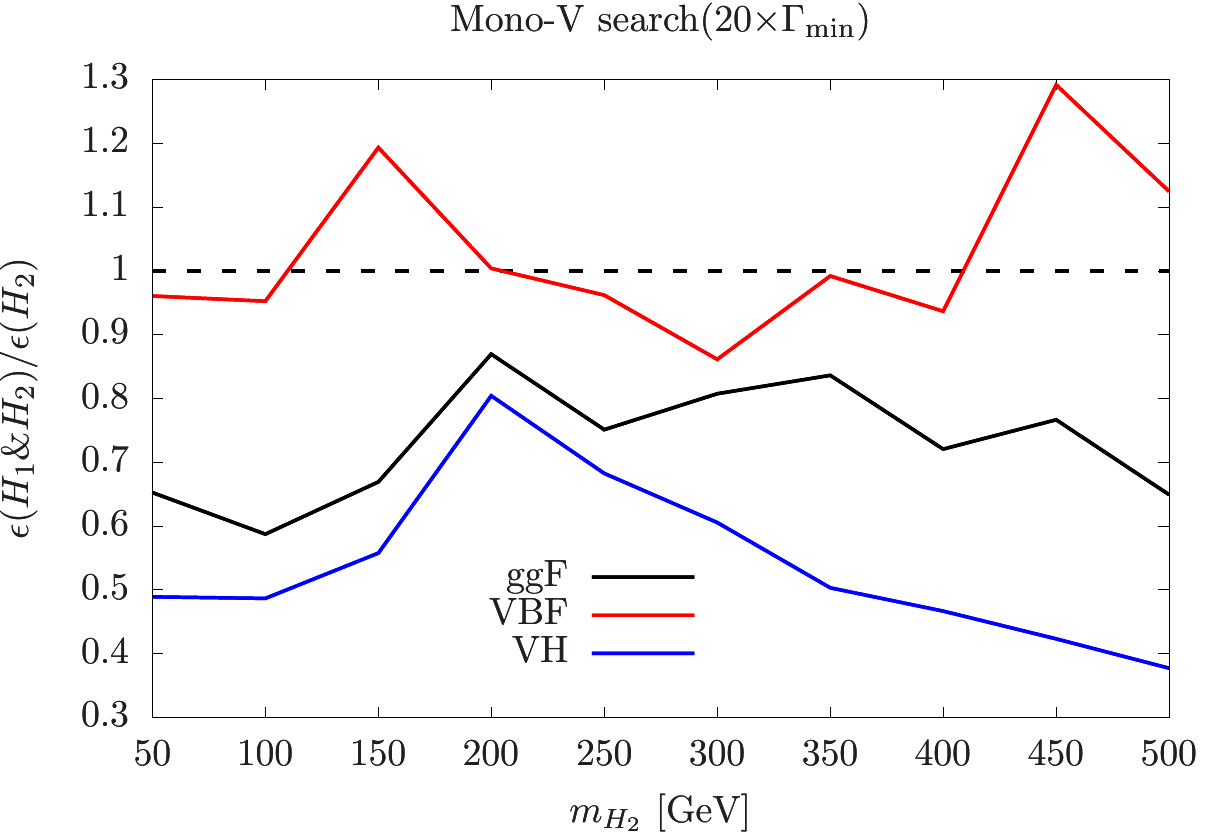}
 \caption{\label{fig:eff2} The cut efficiency ratios for mono-jet search (upper panels) and mono-V search (lower panels) in different production processes. 
Left panels: The decay width of $H_2$ is $\Gamma_{\min}$. Right panels: A relatively large decay width of $H_2$ is chosen, $20\times \Gamma_{\min}$.  }
\end{figure}

Three production processes are analyzed individually. The signal efficiency of mono-jet/mono-V search for a given benchmark point in a production process is defined as the ratio between the number of signal events in mono-jet/mono-V SR and the total number of simulated events. In order to show the influence of the interference effect to signal efficiency, we plot the efficiency ratios between signals with and without $H_1$, i.e. $\epsilon(H_1\&H_2)/\epsilon(H_2)$, for different production processes with varying $\Gamma_{H_2}$ and $m_{H_2}$ in Fig.~\ref{fig:eff2}.

As we can see from the figure, because the interference effect always enhances the production rate in 
low energy scale region, the efficiency ratios are smaller than one in most cases.  
In mono-jet search, when $m_{H_2} < 2 m_\chi$, the amplitudes of the Feynman diagrams with $H_1$ and $H_2$ propagators are of similar size, so the destructive interference effect is significant, especially in the high energy scale region. When the $H_2$ is not so heavy and has small decay width, the amplitude of $H_2$ becomes dominant once it goes on-shell, $m_{H_2} > 2 m_\chi$. Then, the interference effect is negligible. The interference effect is more important for scenarios with wider $H_2$ decay width. For $m_{H_2} \in [50,500]$ GeV and $\Gamma(H_2) = 20\times \Gamma_{\min}$ the interference effect can change the signal efficiency by around 20\% in ggF and VBF production processes. The VH process is most sensitive to the interference effect because of the higher energy scale, where the efficiency ratio decrease quickly as increasing either $m_{H_2}$ or $\Gamma(H_2)$. For $m_{H_2}=500$ GeV and $\Gamma(H_2)=20\times \Gamma_{\min}$ , the $\epsilon(H_1\&H_2)/\epsilon(H_2)$ in VH process can be as low as $\sim 50\%$, while it is $\sim 80\%$ for $\Gamma(H_2)=\Gamma_{\min}$ or $m_{H_2}=200$ GeV. 
Comparing to the mono-jet search, the typical signal efficiency in the mono-V search is more than one order of magnitude smaller, thus larger fluctuations of efficiency ratios due to lower statistics are observed. 
For ggF and VH processes, the main trends of $\epsilon(H_1\&H_2)/\epsilon(H_2)$ with varying $m_{H_2}$ and $\Gamma(H_2)$ follow that in the mono-jet search, with an overall shifting downward. 
On the other hand, the efficiency ratio of VBF process is quite insensitive to the interference effects in the mono-V search, irrespective of the $H_2$ decay width.  

\begin{figure}[htb] \centering
\includegraphics[width=0.47\textwidth]{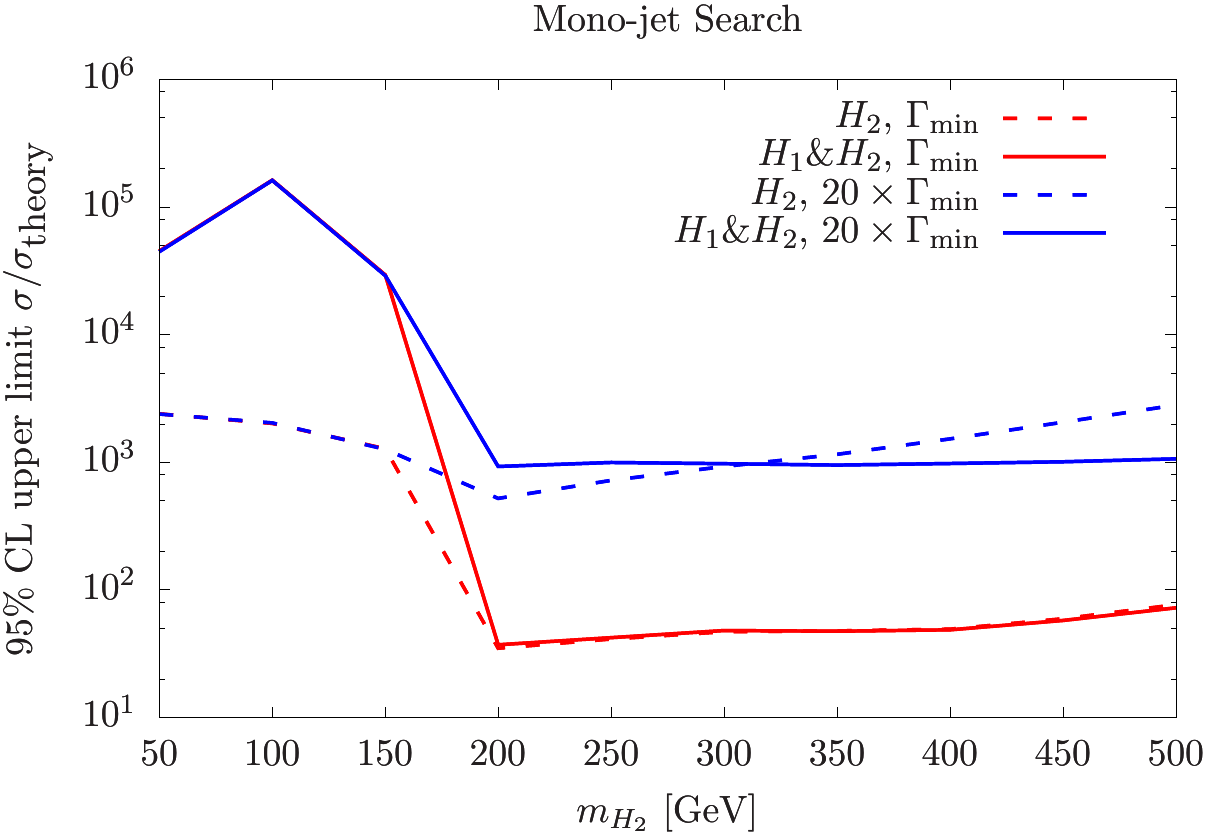}
\includegraphics[width=0.47\textwidth]{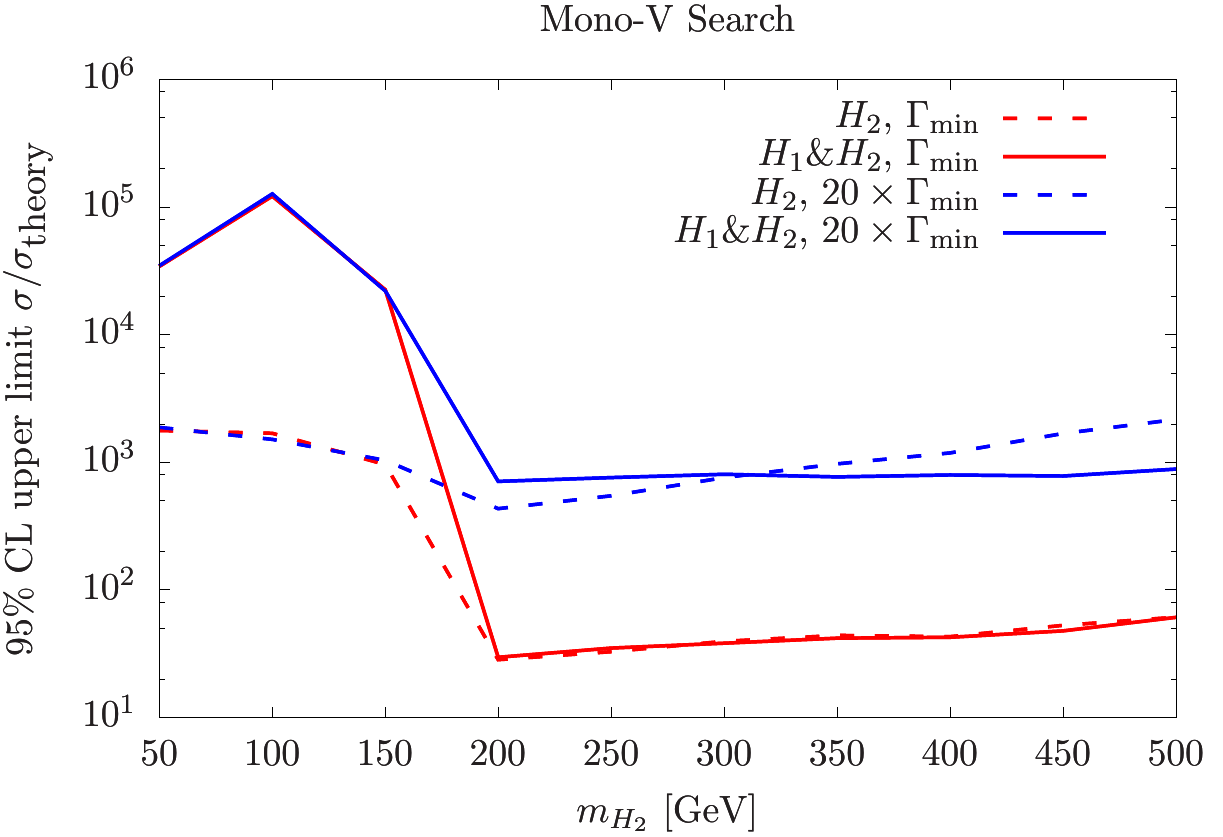}
 \caption{\label{fig:excl2} The CMS exclusion limits on our simplified models. Left: upper limit from mono-jet search. Right: upper limit from mono-V search.  }
\end{figure}

Known the production cross sections and the signal efficiencies, we now able to calculate the CMS search constraints on our models. For a given parameter point, the number of signal events in the mono-jet/mono-V SR is calculated by $\mathcal{L} \times \sigma_i \times \epsilon^{\text{mono-jet/mono-V}}_i$, where $\mathcal{L}=12.9$ fb$^{-1}$ is integrated luminosity and $i$ indicates the production process, ggF, VBF or VH.  The contributions from all three production processes are added up which will be compared with the $N^{\text{upper}}_{\text{mono-jet}}$ in mono-jet search and $N^{\text{upper}}_{\text{mono-V}}$ in mono-V search. The limit on the signal strength 
\begin{align}
\mu_{\text{mono-jet/mono-V}} = \frac{\sigma}{\sigma_{\text{theory}}} = \frac{N^{\text{upper}}_{\text{mono-jet/mono-V}}}{\sum_{i={\text{ggF, VBF, VH}}} \mathcal{L} \times \sigma_i \times \epsilon^{\text{mono-jet/mono-V}}_i  } 
\end{align}
for $H_1\&H_2$ and $H_2$ scenarios are plotted in Fig.~\ref{fig:excl2}.

From Fig.~\ref{fig:excl2} we can observe  that the features of the exclusion bounds are approximately described
by the inverse of the production cross sections. In the light $H_2$ region $m_{H_2} < 2m_\chi$, the reduction 
of cross section due to destructive interference leads to very weak bound in the $H_1\&H_2$ scenario. 
The bounds become much more stringent when $m_{H_2} \gtrsim 2m_\chi$ because of the resonant 
enhancement, especially for narrow decay width of $H_2$. 
However, the interference effect on signal efficiencies also play non-negligible roles in the exclusion bounds. 
As we have discussed for Fig.~\ref{fig:xsec}, the interference effect on cross section leads to smaller cross 
section when $m_{H_2} \in (2 m_{\chi\chi}, 270~\text{GeV})$ and larger cross section when $m_{H_2} > 270$ GeV. The reduction of signal efficiency from interference effect will the enlarge the difference in search sensitivities for $m_{H_2} \in (2 m_{\chi\chi}, 270~\text{GeV})$ and shrink it for $m_{H_2} > 270$ GeV, as can be seen clearly from the solid and dashed blue curves in Fig.~\ref{fig:excl2}. 
Among two searches, the mono-V search has slightly better sensitivity than the mono-jet search. Both of them are indicating that the signal cross section in our model is at least one order of magnitude below the current reach. This is mainly because of the suppression factor of $\sin^2 2\alpha$ in all DM production cross sections. Much larger data set or/and higher hadron collision energy are expected to probe our models. 

\begin{table}[h!]\centering
 \begin{tabular}{c|c|c|c|c|c|c}
   & \multicolumn{3}{c|}{Mono-jet SR} & \multicolumn{3}{c}{Mono-V SR}  \\ \cline{2-7}
    & ggF & VBF & VH & ggF & VBF & VH  \\ \hline
  $H_2, \Gamma_{\min}$ & 194.4 & 22.3 & 2.9                & 7.8 & 1.2 & 1.4 \\
  $H_1\&H_2, \Gamma_{\min}$ & 197.0 & 22.7 & 3.2      & 7.7 & 1.3 & 1.5 \\
  $H_2, 20\times\Gamma_{\min}$ & 6.2 & 0.82 & 0.092   & 0.28 & 0.049 & 0.043 \\
  $H_1\&H_2, 20\times\Gamma_{\min}$ & 9.2 & 1.5 & 0.28  & 0.36 & 0.094 & 0.11 \\
 \end{tabular}
 \caption{\label{tab:bench} The number of events of different production processes in mono-jet SR and mono-V SR for each signal process with $m_{H_2}=400$ GeV at 12.9 fb$^{-1}$ 13 TeV LHC. }
\end{table}

The composition of the DM signal in the mono-jet SR and the mono-V SR in terms of three production processes 
for the benchmark point with $m_{H_2} = 400$ GeV are provided in Table~\ref{tab:bench}. 
For mono-jet search, the ggF is always the most dominant process, the composition of which is around one order 
of magnitude larger than that of VBF and around two orders of magnitude larger than that of VH process. 
The VH becomes much more important in the mono-V search, whose composition is only a few times smaller 
than that of ggF. Note that in mono-V search, there are still large contributions from ggF due the mis-tagging of 
boosted vector boson jet. We would also like to point out that the interference effect tends to increase 
the composition of VH in both SRs, especially for the scenario with large $H_2$ decay width.

\section{Conclusions}
\label{sec:conl}

In this paper, we have considered the collider phenomenology of a gauge invariant and renormalizable model 
for singlet fermion DM with Higgs portal. In this model, there appear two neutral scalar bosons formed by 
the mixing of the singlet and the doublet Higgs bosons that could mediate the DM production.
In certain kinematic regions, their interference can affect the signal in either destructive or constructive manners.
This leads to very important applications to the DM searches at the LHC which have been largely ignored in 
the previous study except in Ref.~\cite{Talk:ko_talks,Baek:2015lna}. 

Due to the minus sign in the scalar mixing matrix (\ref{eq:hmix}), the DM production rate is enhanced in 
the kinematic region where $m_{H_1}< m_{\chi\chi} <m_{H_2}$ and suppressed in the region of 
$m_{\chi\chi}<m_{H_1}$ or  $m_{\chi\chi}>m_{H_2}$, thus affecting  both DM production cross sections and final 
state distributions.  The cross section will be reduced substantially when the $m_{H_2}<2 m_{\chi}$, 
thereby rendering the collider search for DM less effective than naively expected from the simplified model 
without the SM Higgs-like $H_1$  propagator. 
For $m_{H_2}>2 m_{\chi}$, the interference effect in cross section changes from destructive to constructive 
one with increasing the $H_2$ mass.    
Assuming $m_{H_1}<2m_{\chi}$, such that the SM Higgs invisible decay is kinematically forbidden, 
the interference effect always enhances the signal events in the low $m_{\chi\chi}$ region. 
Since the frequency and the energy of the jet radiation are proportional to $m_{\chi\chi}$, lower jet multiplicity 
and less energetic jets in the final state are obtained. 
We also find the interference effect becomes more important for larger decay width of $H_2$. 

The CMS search for DM in final states with either an energetic jet or a boosted hadronically decaying vector 
boson has been applied to scenarios with and without $H_1$. The interference effect will dramatically 
reduce the sensitivity in the parameter region $m_{H_2}<2 m_{\chi}$ while the suppression is only mild when 
$m_{H_2}$ becomes slightly higher than $2 m_{\chi}$. The sensitivity can even be enhanced when $H_2$ is 
significantly higher than $2 m_{\chi}$.  
In our model setup the production cross section of DM signals is more than one order of magnitude below the 
current LHC search sensitivity, mainly because of the small factor $\sin^2 2\alpha$ in production cross section.

\acknowledgments
This work is supported in part by National Research Foundation of Korea (NRF) Research Grant NRF-2015R1A2A1A05001869 (PK, JL), and by the NRF grant funded by the Korea government (MSIP) 
(No. 2009-0083526) through Korea Neutrino Research Center at Seoul National University (PK).

\bibliography{HiggsDM}

\end{document}